\documentclass{llncs}
\usepackage{hyperref}
\usepackage{graphicx}
\usepackage{xspace}
\usepackage{amsmath}  
\usepackage[subtle,mathspacing=normal]{savetrees}
\usepackage{multirow}
\usepackage{xcolor}

\newcommand{\fullonly}[1]{}
\newcommand{\shortonly}[1]{#1}

\newcommand{\lncsonly}[1]{#1}
\newcommand{\articleonly}[1]{}
\newcommand{\acmonly}[1]{}
\newcommand{\svonly}[1]{}

\newcommand{\mycomment}[1]{}

\acmonly{
\copyrightyear{2017} 
\acmYear{2017} 
\setcopyright{acmcopyright}
\acmConference{SACMAT '17}{June 21--23, 2017}{Indianapolis, IN, USA}
\acmPrice{15.00}
\acmDOI{http://dx.doi.org/10.1145/3078861.3078878}
\acmISBN{978-1-4503-4702-0/17/06}
}

\articleonly{}
\lncsonly{}
\acmonly{}

\newcommand{\tuple}[1]{\langle #1\rangle}
\newcommand{\mean}[1]{\left[ \! \left[ #1 \right]\! \right]}

\newcommand{\intersect}{\cap}

\newcommand{\om}{{\it OM}}

\newcommand{\Act}{{\it Act}}
\newcommand{\nav}{{\rm nav}}
\newcommand{\val}{{\it val}}

\newcommand{\op}{{\it op}}
\newcommand{\Rules}{{\it Rules}}
\newcommand{\spa}{E}

\newcommand{\ind}{\hspace*{0.7em}}

\newcommand{\union}{\cup}

\newcounter{lnum}

\newcommand{\com}[1]{{\it #1}}

\newcommand{\function}{{\bf function}}

\newcommand{\foreachloop}{{\bf for each}}

\newcommand{\return}{{\bf return}}

\begin{document}

\newcommand{\thanksText}{This material is based on work supported by FY24-25 CSU RSCA and COS DCI Grants.
}

\title{An Approach for Handling Missing Attribute Values
in Attribute-Based Access Control Policy Mining\lncsonly{\thanks{\thanksText}}}

\articleonly{\author{Thang Bui and Scott~D.~Stoller and Jiajie Li\\
    Department of Computer Science, Stony Brook University, USA}}

\lncsonly{\author{Thang Bui \and Elliot Shabram \and Anthony Matricia}
  \institute{School of Computing and Design, California State University, Monterey Bay, USA}}

\acmonly{
\author{Thang Bui}
\affiliation{
\institution{Stony Brook University}}
\email{thang.bui@stonybrook.edu}
\author{Scott~D.~Stoller}
\affiliation{
\institution{Stony Brook University}}
\email{stoller@cs.stonybrook.edu}
\author{Jiajie Li}
\affiliation{
\institution{Stony Brook University}}
\email{jiajie.li@stonybrook.edu}
}

\svonly{\author{Thang Bui \and Scott~D.~Stoller \and Jiajie Li}

\institute{T. Bui \and S. D. Stoller \and J. Li \at Stony Brook University, USA\\ \email{stoller@cs.stonybrook.edu}}

\date{Received: date / Accepted: date}}

\newcommand{\abstracttext}{ %
    Attribute-Based Access Control (ABAC) enables highly expressive and flexible access decisions by considering a wide range of contextual attributes. ABAC policies use logical expressions that combine these attributes, allowing for precise and context-aware control. Algorithms that mine ABAC policies from legacy access control systems can significantly reduce the costs associated with migrating to ABAC. However, a major challenge in this process is handling incomplete entity information, where some attribute values are missing.
 
    This paper introduces an approach that enhances the policy mining process by predicting or inferring missing attribute values. This is accomplished by employing a contextual clustering technique that groups entities according to their known attributes, which are then used to analyze and refine authorization decisions. By effectively managing incomplete data, 
    our approach provides security administrators with a valuable tool to improve their attribute data and ensure a smoother, more efficient transition to ABAC.
}

\acmonly{
\begin{abstract}
\abstracttext
\end{abstract}}




\acmonly{\thanks{\thanksText}}

\maketitle

\lncsonly{
\begin{abstract}
\abstracttext
\end{abstract}}
\articleonly{
\begin{abstract}
\abstracttext
\end{abstract}}
\svonly{
\begin{abstract}
\abstracttext
\end{abstract}}


\section{Introduction}
\label{sec:intro}

In attribute-based access control (ABAC), access control policies are expressed in terms of a wide range of attributes related to users, resources, actions, and the environment.  This increases expressiveness and often allows more natural policies.  ABAC is becoming increasingly important, as policies become more dynamic and complex. This is reflected in the widespread transition from access control lists (ACLs) to role-based access control (RBAC), and more recently in the ongoing transition from ACLs and RBAC to ABAC. ABAC policy models allow concise policies and promise long-term cost savings through reduced management effort.


Developing ABAC policies can be costly, which can hinder adoption. However, ABAC {\em policy mining} algorithms have the potential to significantly reduce this cost by automatically generating draft ABAC policies from existing lower-level data, such as access control lists (ACLs) or access logs. There is substantial research on ABAC policy mining, as surveyed in \cite{das2018}. 

The fundamental ABAC policy mining problem is: Given information about the attributes of entities (users and resources) in the system and the currently granted permissions, find an ABAC policy that grants the same permissions using concise, high-level ABAC rules.  Several papers consider a variant of this problem where the information about permissions is incomplete \cite{bui18mining,sparselogs2018,iyer2020active,law20fastlas,xu14miningABAClogs}.  However, most existing works on ABAC policy mining assume that the attribute information is complete, i.e., all attributes of all entities have known values. Unfortunately, in most real-world data, some attribute values are missing. Bui et al. \cite{bui2020} allow attributes to be unknown and address the challenge by employing a three-valued logic learning formula to mine policies without attempting to replace the missing values. While this approach avoids the need for imputation, it has limitations. It does not provide security administrators with potential values for the missing attributes, which could be valuable for improving data quality. Additionally, this method requires extensions to the policy language, which could complicate the migration process and limit compatibility with other policy mining approaches.

{\em This paper introduces an algorithm to address the challenge of missing attribute values by predicting or inferring them.} The motivation for our method is the observation that users and resources with similar characteristics often display correlations in their authorizations or permissions. Specifically, users within the same functional group tend to have similar access rights to certain resources, and this correlation can be effectively leveraged to infer missing attribute values with greater accuracy.

Our algorithm employs a contextual clustering technique that groups entities (users and resources) based on their known attributes. Once the clustering is complete, a regression algorithm is then used to learn the correlations in authorizations among these groups. By analyzing these correlations, we can make informed predictions about the missing attribute values. This step leverages the assumption that entities within the same cluster are likely to have similar access permissions, allowing us to fill in the gaps in attribute data effectively.

This approach enhances the quality of the mined policies and supports security administrators by providing insights into potential attribute values, facilitating better decision-making. It also provides valuable insights for the policy mining process by identifying important features learned through our analysis, which could be further extended in future work to develop efficient ABAC policy mining algorithms based on these features. This extension would allow our algorithm not only to predict missing values but also to generate high-quality ABAC policies.

We evaluate our algorithm through two case studies, and the results demonstrate its effectiveness in accurately predicting missing attribute values. Our method contributes to a smoother and more efficient transition to ABAC, reducing the likelihood of errors during the migration process and improving overall access control management.

\section{Policy Language}
\label{sec:language}

We adopt Xu et al.'s  \cite{xu15miningABACShort} ABAC policy language, which is also utilized in Bui et al.'s  \cite{bui2020} work mentioned in Section \ref{sec:intro}. We chose this policy language because it is more expressive than other policy languages that have been used in work on ABAC mining. We describe the language briefly and refer the reader to \cite{xu15miningABACShort} for details.  

An {\em ABAC policy} is a tuple $\pi=\tuple{\om, \Act, \Rules}$, where $\om$ is an object model, $\Act$ is a set of actions, and $\Rules$ is a set of rules.

An {\em object model} is a collection of objects that represent both users and resources within the system. Each object is defined by a set of attribute-value pairs, which describe the properties of a user or resource. Attributes can be single-valued, holding atomic values such as strings or Boolean values, or multi-valued, where the attribute contains a set of atomic values. These attribute-value pairs form the basis for representing and distinguishing between different users and resources in the model. For example, in a healthcare policy, a doctor object (user) might be defined as follows: $\tuple{{\rm\textsf id=doc1024}, {\rm specialties=\{cardiology, electrophysiology\}}, {\rm isTrainee=False}}$. This representation captures the doctor's unique identifier, areas of specialization, and trainee status, effectively modeling the relevant attributes within the system.

A {\em condition} is a set, interpreted as a conjunction, of atomic conditions.  
An {\em atomic condition} is a tuple $\tuple{attr, \op, \val}$, where $attr$ is an attribute, $\op$ is an operator, either ``in'' ($\in$) or ``contains'' ($\ni$), and $\val$ is a constant value, either an atomic value or a set of atomic values.  
For example, a doctor object $o$ satisfies $\tuple{{\rm specialties}, {\rm contains}, {\rm dermatology}}$ ($\rm specialties \ni \rm dermatology$) if the set of values obtained from the attribute specialties of  $o$ contains dermatology. 

A {\em constraint} is a set, interpreted as a conjunction, of atomic constraints.  Informally, an atomic constraint expresses a relationship between the requesting user and the requested resource, by relating the values of specified attributes from each of them.  An {\em atomic constraint} is a tuple $\tuple{attr_u, \op, attr_r}$, where $attr_u$ and $attr_r$ are specified user attribute and resource attribute, respectively, and $\op$ is one of the following four operators: ``equal'' ($=$), ``in'' ($\in$), ``contains'' ($\ni$), ``supseteq'' ($\supseteq$). For example, a doctor (user) $u$ and a consultant (resource) $r$ satisfy $\tuple{{\rm specialties}, {\rm supseteq}, {\rm topics}}$ ($\rm specialties \supseteq \rm topics$) if the set $u$.specialties is a superset or equals the set $r$.topics. This implies that the doctor specializes in all the topics relevant to the consultation.


A {\em rule} is a tuple $\langle {\it userCondition},{\it resourceCondition},$ ${\it constraint}, {\it actions}\rangle$, where {\it userCondition} and {\it resourceCondition} are conditions, {\it constraint} is a constraint, {\it actions} is a set of actions.
For a rule $\rho=\tuple{uc, rc, c, A}$, let $uc$, $rc$, $c$, and $A \subseteq Act$ be the user condition, resource condition, constraint, and set of actions of $\rho$, respectively. 

An {\em entitlement} is represented as a tuple $\tuple{{\it u}, {\it r}, {\it a}}$, indicating that user {\it u} is authorized to perform action {\it a} on resource {\it r}.

An object $o$ {\em satisfies} an atomic condition $c=\tuple{attr, \op, \val}$, denoted $o\models c$, if $(\op={\rm in} \land o.attr \in \val) \lor (\op={\rm contains} \land o.a \ni \val)$.\shortonly{  Objects $o_1$ and $o_2$ {\em satisfy} an atomic constraint $c=\tuple{attr_u, \op, attr_r}$, denoted $\tuple{o_1,o_2} \models c$, is defined in a similar way.}\fullonly{  The {\em meaning} of a condition $c$ relative to a class $C$, denoted $\mean{c}_C$ is the set of instances of $C$ (in the implicitly given object model) that satisfy $c$.  A condition $c$ {\em characterizes} a set $O$ of objects of class $C$ if $O$ is the meaning of $c$ relative to $C$.
Objects $o_1$ and $o_2$ {\em satisfy} an atomic constraint $c=\tuple{p_1, \op, p_2}$, denoted $\tuple{o_1,o_2} \models c$, if $(\op={\rm equal} \land \nav(o_1,p_1) = \nav(o_2,p_2)) \lor (\op={\rm in} \land \nav(o_1,p_1) \in \nav(o_2,p_2)) \lor (\op={\rm contains} \land \nav(o_1,p_1) \ni \nav(o_2,p_2)) \lor (\op={\rm supseteq} \land \nav(o_1,p_1) \supseteq \nav(o_2,p_2))$.
}
An entitlement $\tuple{u, r, a}$ {\em satisfies} a rule $\rho=\langle uc, $ $rc, c, A\rangle$, denoted $\tuple{u, r, a} \models \rho$, if
$u\models uc \land  r\models rc  \land \tuple{u,r}\models c \land a \in A$.  The {\em meaning} of a rule $\rho$, denoted $\mean{\rho}$, is the set of entitlements that satisfy it.
The {\em meaning} of a ABAC
policy $\pi$, denoted $\mean{\pi}$, is the union of the meanings of its rules.



\section{Problem Definition}
\label{sec:problem}


We extend Bui et al.'s  \cite{bui2020} policy mining problem with unknown attribute values by shifting the focus from rule mining to predicting the values of these missing attributes. Given a set $\spa_0$ of entitlements and an object model $\om$, which may contain missing attribute values, our goal is to predict the missing attribute values with one of three confidence levels: High, Medium, or Not Enough Information (NEI). When there is insufficient information to make a confident prediction, it may indicate that the missing values do not significantly impact the entitlements or authorizations and that these attributes might not be necessary for constructing rules in the policy mining process. Note that the set of entitlements represents an ACL policy. Our problem definition could also work with RBAC policies by deriving the set of entitlements from the RBAC policy and using that entitlement set as input.
\section{Algorithm}
\label{sec:algo}

Our algorithm consists of two phases. In the first phase, we group users and resources by applying a contextual clustering method that leverages their known attributes. In the second phase, we employ a regression algorithm to learn the correlations within the input entitlements across these user and resource groups. We formulate the features as atomic conditions and constraints, allowing them to effectively capture the attribute information relevant to each group. This process identifies and ranks the most significant features influencing these correlations, which are then used to infer the missing attribute values. Each inferred value is assigned a confidence level based on the importance of the features involved.

\subsection{Phase 1: Clustering Users and Resources}
\label{sec:algo:phase1}

\noindent In this phase, we first group users and resources based on their sets of active attributes. The set of active attributes of an object consists of those attributes that do not have a {\tt NULL} value. According to Bui et al. \cite{bui2020}, {\tt NULL} is a special value indicating that the attribute is not applicable to the object. For instance, a Student object would have the attribute isChair = {\tt NULL} since isChair is only relevant for Faculty objects to specify whether a faculty member is the chair of their department. This is distinct from a missing value. This approach allows us to effectively identify and cluster objects with common traits, as each distinct set of active attributes will form a unique group, ensuring that objects with similar characteristics are grouped together.

To further refine the groups, we use an object similarity metric that measures the overlap of attribute values between two objects based on Jaccard similarity. Each categorical attribute value is treated as a singleton set. For two objects, $a$ and $b$, the similarity of the $i$-th attribute is defined as as $J_i(a, b) = |A_i\intersect B_i| \,/\, |A_i \union B_i|$, where $A_i$ and $B_i$ are the sets of values for the $i$-th attribute in objects $a$ and $b$. When attribute values are missing, we assign a default similarity of 0.5 to avoid bias and preserve balance when hadnling incomplete data. Users can also specify the importance of each attribute by assigning weights in the similarity measure between two objects, $a$ and $b$, defined as follows: $sim(a,b) = (\sum_{i=1}^{n} w_i \cdot J_i(a, b)) \,/\, (\sum_{i=1}^n w_i)$, where $w_i$ represents the importance weight of the $i$-th attribute, and $n$ is the total number of attributes in the considered set of active attributes. 

For each object within a group defined by a specific set of active attributes, we compute its similarity with all other objects in the group and average the results. We introduce a threshold parameter, ST (Similarity Threshold), to further refine the groups. Objects with an average similarity below ST are separated into new groups, and this process is iteratively applied to the newly created groups until no further divisions are possible. The final refined groups are then utilized in the next phase of the algorithm.

\subsection{Phase 2: Group Analyzing and Predicting Missing Values}
\label{sec:algo:phase2}

A {\em feature} is an atomic condition (on a user or resource) or atomic constraint.  We define a mapping from {\em feature vectors} to Boolean labels: given a tuple $\tuple{u,r,a}$, we create a feature vector (i.e., a vector of the Boolean values of features evaluated for subject $u$ and resource $r$) and map it to true if the tuple is permitted (i.e., is in $\spa_0$) and to false otherwise.  We represent Booleans as integers: 0 for false, and 1 for true. We use linear regression to learn this classification (labeling) of feature vectors corresponding to the relevant groups of users and resources. 

\begin{figure}[tbp]
\begin{tabular}[t]{@{}l@{}}
\function\ predictMissingUserAttr($u, attr_m, E_0, OM)$ \\
\ind //\com{Step 1: Retrieve relevant groups} \\
\ind $groups_{ura} = \emptyset$ //\com{set of (userGroup, resourceGroup, action) tuples}\\
\ind $E_u$ = set containing entitlements which involves $u$\\
\ind \foreachloop\ $(u_e, r_e, a_e) \in E_u$\\
\ind \ind $groups_{ura}$.add((group($u_e$), group($r_e$), $a_e$))\\

\\

\ind //\com{Step 2: Predict missing attribute value} \\
\ind $prediction = \emptyset$ \\ 
\ind \foreachloop\ $g_u, g_r, a$ in $groups_{ura}$ \\
\ind \ind  $learningData$ consists of feature vectors that involve users in $g_u$ and resources in $g_r$, \\
\ind \ind \ind with labels determined by the permissions associated with action $a$ in $E_0$\\ 
\ind \ind //\com{Use linear regression to learn the set of important features}\\ 
\ind \ind $importantFeatures = $ learnImportantFeatures($learningData$) \\
\ind \ind $predictions$.add(predictMissingValues($importantFeatures$, $attr_m$, $OM$))\\
\ind \return\ combinePredictions($predictions$)\\

\end{tabular}
\caption{Algorithm for predicting a user missing attribute value.}
\label{fig:phase2-alg}
\end{figure}

The pseudocode for the algorithm to predict a missing value of an attribute $attr_m$ for user $u$ is shown in Figure \ref{fig:phase2-alg}. In the pseudocode, the function group($o$) returns the group of object $o$ as determined in Phase 1. The algorithm can be similarly applied to predict missing attribute values for resources, following the same procedure. The algorithm consists of two main steps. 
In the first step, the algorithm iterates over the set of entitlements that grant permissions to user $u$ and retrieves a set of tuples consisting of the user group, the resource group, and the action associated with each entitlement. Note that the user group in all these tuples is the same, which is group($u$). This step is crucial for identifying the relevant resource groups that user $u$ has permissions for. It also aligns with the motivation discussed in Section \ref{sec:intro} that entities within the same cluster are likely to have similar access permissions. 

In the second step, the algorithm examines the correlations in permissions among entities within the identified user group, resource group, and action tuples to determine a set of important features. These features are then used to predict the missing attribute values. The learning data consists of feature vectors associated with all users and resources within the considered groups, where the label is set to true (permitted) if the user is allowed to perform the considered action according to $E_0$, and false otherwise, as explained above. It is important to note that feature vectors for users and resources with missing attribute values are excluded from the learning data to ensure the accuracy of the feature extraction and prediction process. 

Function learnImportantFeatures uses linear regression to learn the correlations between attributes and permissions. The output is a ranked list of features based on their learned coefficients. Linear regression is used because it provides a simple, interpretable model that directly shows the relationship between attributes and permissions, allowing us to easily identify and rank important features while being computationally efficient for large datasets. Its straightforward outputs make it ideal for understanding and justifying access decisions based on attribute correlations.

The function predictMissingValues predicts the missing values of attribute $attr_m$ using the ranked list of features obtained from the previous step. We introduce a parameter, NTCF (short for ``number of top confident features''), to determine the confidence level of our predictions. NTCF is defined as a tuple $\tuple{numHigh, numMed}$. A missing value predicted from a feature ranked within the top $numHigh$ features is assigned a high confidence level, while a prediction based on features ranked between $numHigh$ and $numMed$ is assigned medium confidence. For each top-ranked feature within $numMed$, denoted as $f$, the algorithm checks whether $attr_m$ is part of $f$, noting that each feature is either an atomic condition or an atomic constraint. If $f$ is an atomic condition $\tuple{attr_m, \op, \val}$, we predict $\val$ as the missing value. If $f$ is an atomic constraint $\tuple{attr_m, \op, attr_r}$, we gather all values of $attr_r$ for resources in the relevant resource group and, based on the constraint operator $\op$, predict one or multiple values for $attr_m$ from the retrieved set. No predictions are made if $attr_m$ is not present in $f$, and features ranked lower than $numMed$ are not considered. 

After obtaining all predictions with their respective confidence levels for $attr_m$ from the runs with different (user group, resource group, and action) tuples, the combinePredictions function makes the final predictions by selecting the predicted value with the highest confidence level for each distinct value, and it determines the number of values to return based on whether $attr_m$ is a single-valued or set-valued attribute. It then returns a set of final predictions along with their confidence levels. If no predictions are made, the algorithm refrains from predicting any values for $attr_m$ and notifies the administrator, advising them to proceed with caution. The absence of predictions may indicate that the missing values do not significantly affect the entitlements or authorizations, as none of the features involving $attr_m$ were identified as important in the feature learning step. Note that the identified important features offer valuable insights that could be used in policy mining, suggesting a potential extension for future work to integrate feature-driven policy generation and enhance the overall quality of mined policies.

\subsection{Example}
\label{sec:algo:example}

\begin{figure}[tbp]
\begin{tabular}[t]{@{}l@{}}
\tt{\# Object model:}\\
\tt{\# Users:}\\
\tt{u1:}$\tuple{{\rm\tt id=csFac1}, {\rm\tt position=faculty}, {\rm\tt\textcolor{red}{\textbf{department=?}}}, {\rm\tt\textcolor{red}{\textbf{coursesTaught=?}}}}$\\
\tt{u2:}$\tuple{{\rm\tt id=csFac2}, {\rm\tt position=faculty}, {\rm\tt department=cs}, {\rm\tt coursesTaught=\{cs601\}}}$\\
\tt{u3:}$\tuple{{\rm\tt id=eeFac1}, {\rm\tt position=faculty}, {\rm\tt department=ee}, {\rm\tt coursesTaught=\{ee101\}}}$\\
\tt{u4:}$\tuple{{\rm\tt id=eeFac2}, {\rm\tt position=faculty}, {\rm\tt department=ee}, {\rm\tt coursesTaught=\{ee601\}}}$\\
\textcolor{gray}{\tt{u5:}$\tuple{{\rm\tt id=csStu1}, {\rm\tt position=student}, {\rm\tt department=cs}, {\rm\tt coursesTaken=\{cs101\}}}$}\\
\textcolor{gray}{\tt{u6:}$\tuple{{\rm\tt id=eeStu1}, {\rm\tt position=student}, {\rm\tt department=ee}, {\rm\tt coursesTaken=\{ee602\}}}$}\\

\tt{\# Resources:}\\
\tt{r1:}$\tuple{{\rm\tt id=cs101gb}, {\rm\tt department=cs}, {\rm\tt course=cs101}, {\rm\tt type=gradebook}}$\\
\tt{r2:}$\tuple{{\rm\tt id=cs601gb}, {\rm\tt department=cs}, {\rm\tt course=cs601}, {\rm\tt type=gradebook}}$\\
\tt{r3:}$\tuple{{\rm\tt id=ee101gb}, {\rm\tt department=ee}, {\rm\tt course=ee101}, {\rm\tt type=gradebook}}$\\
\tt{r4:}$\tuple{{\rm\tt id=ee601gb}, {\rm\tt department=ee}, {\rm\tt course=ee601}, {\rm\tt type=gradebook}}$\\
\tt{r5:}$\tuple{{\rm\tt id=ee602gb}, {\rm\tt department=ee}, {\rm\tt course=ee602}, {\rm\tt type=gradebook}}$\\
\textcolor{gray}{\tt{r6:}$\tuple{{\rm\tt id=csStu1trans}, {\rm\tt department=cs}, {\rm\tt student=csStu1}, {\rm\tt type=transcript}}$}\\
\textcolor{gray}{\tt{r7:}$\tuple{{\rm\tt id=eeStu1trans}, {\rm\tt department=ee}, {\rm\tt student=eeStu1}, {\rm\tt type=transcript}}$}\\
\\

\tt{\# Rules set:}\\
\tt{\# The instructor of a course can update grades in the course's grade book.}\\
$\rho: \tuple{{\rm\tt position} \in {\rm\tt \{faculty\}}, {\rm\tt type} \in {\rm\tt \{gradebook\}},{\rm\tt coursesTaught} \ni {\rm\tt course}, \rm\tt \{modify\}}$\\
\\

\tt{\# Entitlements:}\\
\tt{e1:}$\tuple{{\rm\tt csFac1}, {\rm\tt cs101gb}, {\rm\tt modify}}$\\
\tt{e2:}$\tuple{{\rm\tt csFac2}, {\rm\tt cs601gb}, {\rm\tt modify}}$\\
\tt{e3:}$\tuple{{\rm\tt eeFac1}, {\rm\tt ee101gb}, {\rm\tt modify}}$\\
\tt{e4:}$\tuple{{\rm\tt eeFac2}, {\rm\tt ee601gb}, {\rm\tt modify}}$\\

\end{tabular}
\caption{Sample university policy.}
\label{fig:example-alg}
\end{figure}

We illustrate the algorithm using a fragment of the University case study described in Section \ref{sec:dataset}. The policy, shown in Figure \ref{fig:example-alg}, involves six user objects (four representing faculty members and two representing students) and seven resources (five grade books and two transcripts). For user attributes, {\tt position} specifies whether the user is faculty or a student, {\tt department} specifies their department, {\tt coursesTaught} lists the courses a faculty member teaches, and {\tt coursesTaken} lists the courses a student takes. For resource attributes, {\tt department} specifies the department associated with a grade book or transcript, {\tt course} specifies the course a grade book belongs to, {\tt student} specifies the student linked to a transcript, and {\tt type} indicates whether a resource is a grade book or a transcript. Attributes that do not apply to a particular object are omitted, with their values set to {\tt NULL} as explained in Section \ref{sec:algo:phase1}. In this example, we need to infer two missing attribute values for user {\tt u1}: the {\tt department}, which should be {\tt cs}, and {\tt coursesTaught} attributes , which should be \{{\tt cs101}\}.

To keep this example concise, we consider only one rule $\rho$: the instructor of a course can update grades in the course's grade book. The set of entitlements defines the meaning of $\rho$. As described in Section \ref{sec:problem}, it is important to clarify that the inputs to our algorithm do not include a rule set. Our case studies involve a complete ABAC policy, consisting of both an object model and a rule set. However, for evaluation purposes, we only use the rule set to retrieve the set of input entitlements. A detailed description of our evaluation methodology is provided in Section \ref{sec:eval-method}.

In phase 1, our algorithm clusters users into two groups based on their active attribute sets: the first group, containing users {\tt u1}, {\tt u2}, {\tt u3}, and {\tt u4}, represents faculty members, while the second group, containing users {\tt u5} and {\tt u6}, represents students. Similarly, for resources, the algorithm clusters grade book objects {\tt r1} through {\tt r5} into one group and transcript objects {\tt r6} and {\tt r7} into another. With attribute weights $w_i$ set to 1.0 and the threshold parameter ST set to 0.25, these groups remain as they are, with no smaller subgroups formed. 

In phase 2, since the missing attribute values in this example belong to the same user, both values can be predicted simultaneously. The first step of function predictMissingUserAttr identifies the relevant resource groups for which user {\tt u1} has permissions. In this case, the algorithm considers only the grade book group identified in the previous phase, since {\tt u1} (with ID {\tt csFac1}) has a single entitlement to modify grade book {\tt r1} (with ID {\tt cs101gb}). Furthermore, the other entitlements also specify that faculty members have permission to modify grade books for the courses they teach. 

In the second step, the learning data involving these two groups is created. This learning data consists of a set of features, including those corresponding to the two atomic conditions and the atomic constraint in rule $\rho$. It is expected that these features, {\tt position} $\in$  \{{\tt faculty}\}, {\tt type} $\in$ \{{\tt gradebook}\}, and {\tt coursesTaught} $\ni$ {\tt course},  will also rank as the top three important features in the output of learnImportantFeautres. 

Then, in predictMissingValues, the algorithm predicts the missing values for the {\tt coursesTaught} attribute using the important feature {\tt coursesTaught} $\ni$ {\tt course}. Since {\tt csFac1} has the modify permission on {\tt cs101gb} (given in the entitlement {\tt e1}), the {\tt coursesTaught} attribute for {\tt csFac1} should contain the value of the {\tt course} attribute for {\tt cs101gb}, which is {\tt cs101}. With the NCTF set to $\tuple{3,5}$, this prediction achieves a High confidence level. The other features mentioned are not applicable for predicting the missing value of {\tt coursesTaught}, as they do not involve this attribute. In this example, none of the other top five ranked features involve the {\tt coursesTaught} or {\tt department} attributes. Therefore, the algorithm cannot make any prediction for the missing {\tt department} value, as no considered important features are associated with it. As a result, the algorithm returns NEI (Not Enough Information) for this attribute.

If user {\tt csFac1} teaches more than one course, the algorithm would be able to predict additional courses based on input entitlements for the new courses, and function predictMissingValues would return multiple predictions. Since we have only one prediction for {\tt coursesTaught} in this example, combinePredictions has no effect on the final result.

Overall, this example illustrates how our algorithm predicts missing attribute values by leveraging the observation that users and resources within similar groups often display correlated permissions. Since other faculty members have modification access to the grade books for the courses they teach, a faculty member with a missing {\tt coursesTaught} attribute who shares similar permissions can have their teaching courses inferred based on these permissions. Conversely, because there are no entitlements based on the department of a faculty member, our algorithm refrains from predicting a value for the missing {\tt department} attribute due to insufficient information. This case demonstrates how a missing attribute, when unrelated to entitlements or authorizations, does not impact the overall permissions.


\vspace*{-5pt}

\section{Evaluation}
\label{sec:evaluation}

\vspace*{-5pt}

\subsection{Dataset} 
\label{sec:dataset}
We evaluate our algorithm on the University and Project Management case studies outlined in \cite{xu15miningABACShort}. The University case study is a policy that controls access by students, instructors, teaching assistants, registrar officers, department chairs, and admissions officers to applications (for admission), gradebooks, transcripts, and course schedules. The Project Management case study is a policy that controls access by department managers, project leaders, employees, contractors, auditors, accountants, and planners to budgets, schedules, and tasks associated with projects. More details about these case studies are provided in \cite{xu15miningABACShort}. 

\subsection{Evaluation methodology} 
\label{sec:eval-method}


\begin{figure}[tbp]
\includegraphics[width=1.0\textwidth]{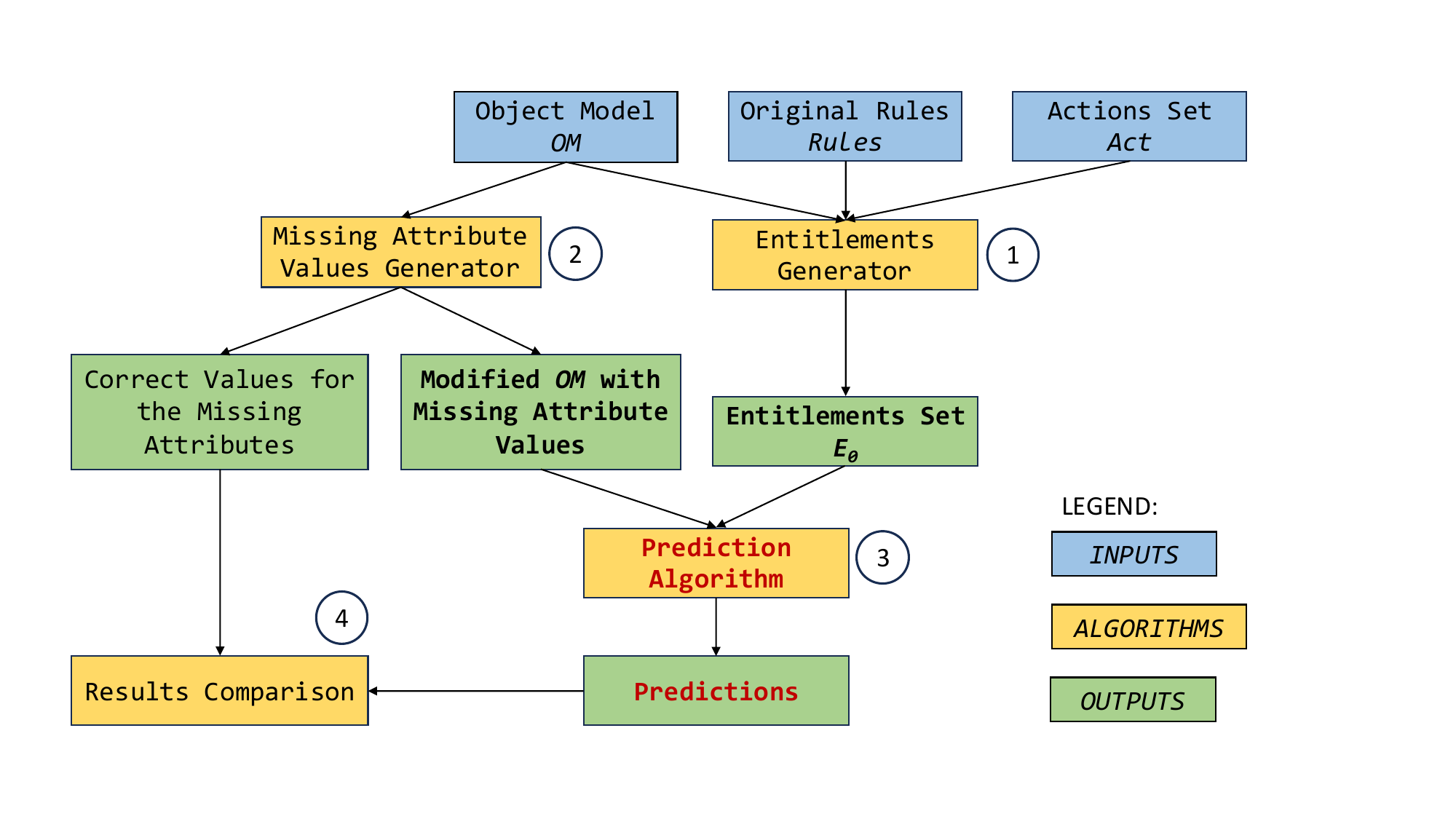}
\caption{Evaluation methodology}
\label{fig:eval-method}
\end{figure}

Our methodology for evaluating the algorithm is depicted in Figure \ref{fig:eval-method}. We start with an ABAC policy $\pi=\tuple{\om, \Act, \Rules}$, compute the set $\mean{\pi}$ of granted entitlements from the object model $\om$ and the set of rules $\Rules$, create from it a set of entitlements $\spa_0$. Then we remove small percentages of attribute values from $\om$. Next, we run our algorithm on $\spa_0$ and the modified $\om$ to predict the missing attribute values and compare the predictions with the original values before their removal. It is important to emphasize that the set of input ABAC policy rules $\Rules$ is only used in the initial step to generate the input entitlements, similar to those found in legacy access control models. Our algorithm operates without access to the ABAC policy rules themselves.

While we use the same dataset as \cite{xu15miningABACShort}, our work addresses a different problem. Their work focuses on mining ABAC rules from complete attribute data, whereas our approach tries to infer missing attribute values in incomplete datasets. Therefore, a direct comparison of the algorithms would not be meaningful, as their algorithm produces a complete set of ABAC rules, while ours predicts missing attribute values.

\subsection{Evaluation Results}  
\label{sec:eval-results}
We ran experiments with three different percentages of missing attribute values: 3\%, 6\%, and 9\%, following the setup used in \cite{bui2020}. The attributes were randomly selected for removal until the desired missing percentage was reached. While \cite{bui2020} allows specifying a set of required or important attributes that are excluded from removal, our approach does not have this restriction. Our algorithm can predict any missing values as long as there is sufficient information, offering better flexibility.

For the algorithm parameters in Phase 1, we set the attribute weights $w_i$ to 1.0 and ST to 0.25. Additional experiments were conducted with ST values of 0.1 and 0.5, which yielded similar results, while setting ST values greater than 0.5 was found to be unreasonable due to its restrictive nature. In Phase 2, we set the NTCF to $\tuple{3,5}$. The choice of setting $numHigh = 3$ was informed by observations from the policy datasets, where rules typically contain no more than three atomic conditions and constraints in total. We set $numMed = 5$ as a heuristic to balance capturing relevant features for medium confidence predictions while avoiding less reliable lower-ranked features. Our algorithm was implemented in Python, and the experiments were conducted on an Apple MacBook Air with an M2 CPU and 8 GB memory.

Experiments were performed with varying policy sizes for each case study, using policies generated from \cite{xu15miningABACShort}. Figure \ref{tab:result-table} summarizes the size of the input entitlements set ($|\spa_0|$), the number of objects and attributes for each policy, as well as the results of our experiments. Each data point represents the average outcome over five runs with random missing attribute inputs for each policy size and missing attributes level. We report two key metrics: coverage and accuracy. Coverage represents the percentage of missing attributes for which our algorithm was able to make predictions. Accuracy measures the correctness of the predictions by comparing them with the original values. Both coverage and accuracy are reported as values between 0.0 and 1.0 instead of percentages. Although the algorithm is designed to predict with high or medium confidence, our experiments only produced high-confidence predictions or none. Therefore, we report a single accuracy value for high-confidence predictions.

\begin{figure*}[tb]
\centering
\begin{tabular}{|c|c|c|c|c|clcl|clcl|clcl|}
\hline
\multirow{3}{*}{Dataset} &
  \multirow{3}{*}{\#objs} &
  \multirow{3}{*}{\#attrs} &
  \multirow{3}{*}{$|\spa_0|$} &
  \multirow{3}{*}{Acc} &
  \multicolumn{4}{c|}{Miss. \%: 3\%} &
  \multicolumn{4}{c|}{Miss. \%: 6\%} &
  \multicolumn{4}{c|}{Miss. \%: 9\%} \\ \cline{6-17} 
 &
   &
   &
   &
   &
  \multicolumn{2}{c|}{\multirow{2}{*}{Cov}} &
  \multicolumn{2}{c|}{\multirow{2}{*}{Time}} &
  \multicolumn{2}{c|}{\multirow{2}{*}{Cov}} &
  \multicolumn{2}{c|}{\multirow{2}{*}{Time}} &
  \multicolumn{2}{c|}{\multirow{2}{*}{Cov}} &
  \multicolumn{2}{c|}{\multirow{2}{*}{Time}} \\
 &
   &
   &
   &
   &
  \multicolumn{2}{c|}{} &
  \multicolumn{2}{c|}{} &
  \multicolumn{2}{c|}{} &
  \multicolumn{2}{c|}{} &
  \multicolumn{2}{c|}{} &
  \multicolumn{2}{c|}{} \\ \hline
university-1 &
  290 &
  797 &
  1588 &
  \textbf{1.0} &
  \multicolumn{2}{c|}{0.85} &
  \multicolumn{2}{c|}{12} &
  \multicolumn{2}{c|}{0.86} &
  \multicolumn{2}{c|}{22} &
  \multicolumn{2}{c|}{0.87} &
  \multicolumn{2}{c|}{23} \\ \hline
university-2 &
  482 &
  1316 &
  3498 &
  \textbf{1.0} &
  \multicolumn{2}{c|}{0.84} &
  \multicolumn{2}{c|}{63} &
  \multicolumn{2}{c|}{0.83} &
  \multicolumn{2}{c|}{107} &
  \multicolumn{2}{c|}{0.83} &
  \multicolumn{2}{c|}{121} \\ \hline
university-3 &
  674 &
  1843 &
  6159 &
  \textbf{1.0} &
  \multicolumn{2}{c|}{0.84} &
  \multicolumn{2}{c|}{236} &
  \multicolumn{2}{c|}{0.88} &
  \multicolumn{2}{c|}{352} &
  \multicolumn{2}{c|}{0.84} &
  \multicolumn{2}{c|}{449} \\ \hline
projmgmt-1 &
  240 &
  976 &
  416 &
  \textbf{1.0} &
  \multicolumn{2}{c|}{0.73} &
  \multicolumn{2}{c|}{15} &
  \multicolumn{2}{c|}{0.71} &
  \multicolumn{2}{c|}{24} &
  \multicolumn{2}{c|}{0.71} &
  \multicolumn{2}{c|}{26} \\ \hline
projmgmt-2 &
  420 &
  1708 &
  728 &
  \textbf{1.0} &
  \multicolumn{2}{c|}{0.69} &
  \multicolumn{2}{c|}{110} &
  \multicolumn{2}{c|}{0.7} &
  \multicolumn{2}{c|}{160} &
  \multicolumn{2}{c|}{0.69} &
  \multicolumn{2}{c|}{200} \\ \hline
projmgmt-3 &
  540 &
  2196 &
  936 &
  \textbf{1.0} &
  \multicolumn{2}{c|}{0.69} &
  \multicolumn{2}{c|}{280} &
  \multicolumn{2}{c|}{0.69} &
  \multicolumn{2}{c|}{440} &
  \multicolumn{2}{c|}{0.71} &
  \multicolumn{2}{c|}{540} \\ \hline
\end{tabular}%
\caption{Experimental results for our algorithm on datasets with different missing percentages (Miss. \%). ``Acc'' is the average accuracy achieved on each policy. ``Acc'' is the average accuracy achieved on each policy.``Time'' is the average running time for each policy, measured in seconds. ``\#object'' and ``\#attrs'' are the average number of objects and the number of attributes for each input policy, respectively.}
\label{tab:result-table}
\end{figure*}

The algorithm consistently achieves 100\% accuracy across all policies and missing attribute percentages, indicating that it made no incorrect predictions when it had sufficient information to do so. {\em This demonstrates the reliability of the algorithm in predicting missing attribute values accurately.}

The coverage results are acceptable, averaging 85\% for the University policies and 70\% for the Project Management policies. While not perfect, these results reflect the algorithm’s cautious approach, making predictions only when sufficient data is available. Notably, the algorithm performs better on the University policies compared to the Project Management policies. This outcome is expected because the Project Management datasets have a smaller set of input entitlements and rules (5 compared to 10 for the University datasets), increasing the likelihood that the randomly chosen missing attributes do not affect any authorization decisions. Across different levels of missing attribute percentages, the algorithm maintains reasonable performance. The coverage remains stable even as the percentage of missing attributes increases, indicating robustness in handling higher levels of missing data. The standard deviation of the reported coverages ranges from 0.01 to 0.06, highlighting consistent performance across multiple runs.

The algorithm’s running time scales with increasing missing attribute percentages. On average, the running time increases by a factor of 1.61 when missing data rises from 3\% to 6\%, and by a factor of 1.17 from 6\% to 9\%. This suggests the algorithm manages increased computational demands reasonably well, even with substantial missing data.


\section{Related Work}
\label{sec:related}


The only prior work on mining of ABAC policies with missing attribute values is \cite{bui2020}, which is discussed in Section \ref{sec:intro}. The authors reduce the core aspect of the ABAC policy mining problem to the problem of learning a formula in Kleene’s three-valued logic.  Three-valued logic allows three truth values: true ($T$), false ($F$), and unknown ($U$).  With three-valued logic, they can assign the truth value $U$ to conditions and constraints involving unknown attribute values, allowing them to handle these uncertainties without the need to predict or impute the missing values. This approach offers the advantage of avoiding potentially inaccurate imputation, which can simplify the initial policy mining process. However, one significant drawback is that it requires extensions to the existing policy language, introducing additional complexity to the system. This can complicate the migration process from legacy systems to ABAC and may reduce compatibility with other policy mining techniques that assume a more traditional binary logic framework.

\subsubsection{Related work on policy mining.}

We also discuss related work on policy mining, as our approach could be integrated with existing algorithms to address missing attribute values. Xu et al. introduced the first algorithm for ABAC policy mining \cite{xu15miningABAC} and later a version for mining ABAC policies from logs \cite{xu14miningABAClogs}. Medvet et al. developed the first evolutionary algorithm for ABAC policy mining \cite{medvet2015}, and Iyer et al. introduced the first algorithm capable of mining policies with both deny and permit rules \cite{iyer2018}. Cotrini et al. reformulated ABAC mining from logs and presented an algorithm using APRIORI-SD, a subgroup discovery technique \cite{sparselogs2018}. They also created a “universal” framework for mining policies across various languages, though their results showed better quality with language-specific algorithms \cite{cotrini2019}. Law et al. presented a scalable inductive logic programming algorithm for learning ABAC rules from logs \cite{law20fastlas}. Bui et al. developed several Relationship-Base Access Control (ReBAC) policy mining algorithms \cite{bui18mining,bui19mining,bui20decision}. The ReBAC policy language they considered extends the ABAC model by formulating ReBAC as an object-oriented extension of ABAC. This approach integrates relationships between entities into the traditional ABAC framework, enabling more expressive access control based on connections among users and resources. Our algorithm could be readily adapted to support this ReBAC policy language, as it builds on similar principles and structure.



\svonly{
\begin{acknowledgements}
\thanksText
\end{acknowledgements}}


%
\acmonly{\bibliographystyle{ACM-Reference-Format}}\lncsonly{\bibliographystyle{splncs04}}\articleonly{\bibliographystyle{alpha}}\svonly{\bibliographystyle{plain}}
\bibliography{references}


\end{document}